\documentclass[5p, twocolumn, longtitle]{elsarticle}

\usepackage{amsmath}   
\usepackage{graphicx}  
\usepackage{verbatim}  
\usepackage{color}     
\usepackage{subfigure} 
\usepackage{hyperref}  
\usepackage{graphics}
\usepackage{bm}
\usepackage{ulem}
\usepackage{lineno}
\usepackage{booktabs}
\usepackage{dcolumn}
\begin{document}

\begin{frontmatter}

\title{First results on nucleon resonance photocouplings from the $\gamma p \to \pi^+\pi^-p$ reaction}

\author[MSU]{E.~Golovatch\corref{cor1}} \ead{golovach@jlab.org} 
\author[JLAB]{V.D.~Burkert} 
\author[JLAB]{D.S.~Carman} 
\author[SCAROLINA]{R.W.~Gothe} 
\author[OHIOU]{K.~Hicks} 
\author[MSU]{B.S.~Ishkhanov} 
\author[JLAB]{V.I.~Mokeev\corref{cor2}} \ead{mokeev@jlab.org} 
\author[JLAB,ASU]{E.~Pasyuk} 

\cortext[cor1]{Corresponding author} 
\cortext[cor2]{Principal corresponding author}

\author[FIU]{S. Adhikari} 
\author[FSU]{Z.~Akbar} 
\author[ODU]{M.J.~Amaryan} 
\author[JLAB]{H.~Avakian} 
\author[SACLAY]{J.~Ball} 
\author[INFNFE]{L. Barion} 
\author[EDINBURGH]{M. Bashkanov}

\author[INFNGE]{M.~Battaglieri} 
\author[ITEP]{I.~Bedlinskiy} 
\author[FU,CMU]{A.S.~Biselli} 

% ...
\address[ASU]{Arizona State University, Tempe, Arizona 85287-1504}
% ...
\address[CMU]{Carnegie Mellon University, Pittsburgh, Pennsylvania 15213, USA}
% ...
\address[SACLAY]{Irfu/SPhN, CEA, Universit\'e Paris-Saclay, 91191 Gif-sur-Yvette, France}
% ...
\address[FU]{Fairfield University, Fairfield CT 06824, USA}
% ...
\address[FIU]{Florida International University, Miami, Florida 33199, USA}
\address[FSU]{Florida State University, Tallahassee, Florida 32306, USA}
% ...
\address[INFNFE]{INFN, Sezione di Ferrara, 44100 Ferrara, Italy}
% ...
\address[INFNGE]{INFN, Sezione di Genova, 16146 Genova, Italy}
% ...
\address[ITEP]{Institute of Theoretical and Experimental Physics, Moscow, 117259, Russia}
% ...
\address[OHIOU]{Ohio University, Athens, Ohio  45701, USA}
\address[ODU]{Old Dominion University, Norfolk, Virginia 23529, USA}
% ...
\address[MSU]{Skobeltsyn Institute of Nuclear Physics and Physics Department, Lomonosov Moscow State
  University, 119234 Moscow, Russia}
\address[SCAROLINA]{University of South Carolina, Columbia, South Carolina 29208}
% ...
\address[JLAB]{Thomas Jefferson National Accelerator Facility, Newport News, Virginia 23606, USA}
% ...
\address[EDINBURGH]{Edinburgh University, Edinburgh EH9 3JZ, United Kingdom}
% ...


\begin{abstract} 
We report the first experimental measurements of the nine 1-fold differential cross sections for the $\gamma p \to \pi^+\pi^-p$ 
reaction, obtained with the CLAS detector at Jefferson Laboratory. The measurements cover the invariant mass range of the 
final state hadrons from 1.6~GeV~$<W<$~2.0~GeV. For the first time the photocouplings of all prominent nucleon resonances 
in this mass range have been extracted from this exclusive channel. Photoproduction of two charged pions is of particular 
importance for the evaluation of the photocouplings for the $\Delta(1620)1/2^-$, $\Delta(1700)3/2^-$, $N(1720)3/2^+$, and
$\Delta(1905)5/2^+$ resonances, which have dominant decays into the $\pi\pi N$ final states rather than the more extensively
studied single meson decay channels. 
\end{abstract} 


\end{frontmatter}

\section{Introduction}
\label{intro} 

Studies of the excitation spectrum of the nucleon and the resonance photocouplings from the experimental data on exclusive 
meson photoproduction represent an important avenue in the exploration of the strong interaction in the non-perturbative
regime~\cite{Az13}. Evaluation of the excited nucleon spectrum within Lattice QCD~\cite{Du12} and continuous QCD
approaches~\cite{Ro11} adds to our understanding of how to relate the experimental results on the $N^*$ spectrum to
the dynamics of strong QCD and its emergence from the QCD Lagrangian. In the past decade, data on exclusive meson
photoproduction off the nucleon have been obtained at CLAS, ELSA, MAMI, GRAAL, and LEPS
\cite{Cr13,Kl17,Be17,Leps,So17,Gu14,Ass03,Kash12,Ann15,Bu16}. The new data include differential cross sections, as well
as single-, double-,  and triple-polarization asymmetries. This wealth of data provides for rigorous constraints on the reaction
amplitudes that are necessary in order to potentially access the amplitudes for two-body final states such as $\pi N$,
$\eta N$, $\eta' N$, $KY$, and $K^*Y$,  to constrain the $\omega p$ and $\phi p$ amplitudes, and to extend the knowledge
on the reaction mechanisms for the double-meson channels $\pi\pi N$ and $\pi\eta N$. 

A global multichannel analysis of these data by the Bonn-Gatchina group~\cite{BnGa12,BnGa14,An16} has provided strong 
evidence for several new baryon states that have been reported in the recent edition of the Review of Particle Properties
(PDG)~\cite{Rpp18}. Strong evidence for the existence of the $N(1710)1/2^+$, $N(1895)1/2^-$, and $N(1900)3/2^+$
resonances has recently become available~\cite{Bu17}. In particular, the CLAS photoproduction data in the $KY$ channels
\cite{Brad06,Brad07,McC09,Dey10} has had a decisive impact on these findings. However, the $\pi^+\pi^-p$
photoproduction data is also sensitive to new baryon states~\cite{Capst,Mo16b}  and offers another complementary channel
to search for such states. Nucleon resonances established in photoproduction can also be observed in exclusive
electroproduction off the proton at different photon virtualities $Q^2$, with $Q^2$-independent masses and hadronic decay
widths. This signature provides strong evidence for the existence of new states. Therefore, combined studies of the
$\pi^+\pi^-p$ photo- and electroproduction data available from CLAS~\cite{Mo16b,Ri03,Is16} can potentially allow for the
validation of the existence of missing baryon states in a nearly model-independent way. These studies have already provided
substantial evidence for the existence of the new $N'(1720)3/2^+$ baryon state~\cite{Mo16b}. 

Furthermore, the $\pi\pi N$ channels of all charge combinations are also a unique source of information on the production of
several well-established resonances with masses above 1.6~GeV. So far, the photocouplings of most $N^*$ and $\Delta^*$
states reported in the PDG were obtained from $\pi N$ and multichannel photoproduction \cite{BnGa12,BnGa14,An16}. The
$\pi\pi N$ photoproduction data analyzed in the mass range above 1.6~GeV include $\pi^0\pi^0p$ data~\cite{So17,Ass03,Kash12},
but do not yet include data on $\pi^+\pi^-p$ cross sections from a proton target. However, the two-body meson-baryon
photoproduction channels have limited sensitivity to many of the resonances with masses above 1.6~GeV, which decay
preferentially into the $\pi\pi N$ final states. Moreover, the $\pi^+\pi^-p$ channel has the largest cross section among
the studied $\pi \pi N$ channels~\cite{Tho05} and is needed to verify the results of other meson-baryon channels
\cite{Mo12,Mo16a}. 

In this paper we present the first data for the nine 1-fold differential $\pi^+\pi^-p$ photoproduction cross sections off the
proton at invariant mass $W$ from 1.6~GeV to 2.0~GeV. These data have allowed us to determine the resonant contributions
from a fit of all measured differential cross sections combined within the framework of the updated Jefferson Lab-Moscow
State University (JM) reaction model~\cite{Mo12,Mo16a,Mo09}. By employing a unitarized Breit-Wigner (BW) ansatz
\cite{Mo12}, the photocouplings of all prominent resonances with masses above 1.6~GeV were extracted from the
$\pi^+\pi^-p$ photoproduction data for the first time.

\section{Experiment}
\label{exp} 

The data were collected using the CEBAF Large Acceptance Spectrometer (CLAS)~\cite{Me03} in Hall~B at the Thomas 
Jefferson National Accelerator Facility during the ``g11a'' data taking period in 2004. The photon beam was produced by an 
unpolarized electron beam of 4.019~GeV energy incident upon a gold-foil radiator with a thickness of $10^{-4}$ radiation
lengths. The photon energies were determined by detecting post-bremsstrahlung electrons in the counters of a tagging 
spectrometer~\cite{tagger}. The tagged-photon energy range was 20-95\% of the electron beam energy. The tagged-photon 
beam impinged on a 40-cm-long LH$_2$ target. The temperature and pressure of this cryotarget were monitored throughout 
the g11a run. The mean calculated density of H$_2$ was 0.0718~g/cm$^3$ with relative fluctuations of about 0.1\% 
\cite{bradnote,willthesis}.

The CLAS spectrometer was based on a six-coil superconducting torus magnet and included a series of detectors situated 
in the six azimuthally symmetric sectors around the beamline. Three regions of drift chambers (DC)~\cite{dcs} allowed for
the tracking of charged reaction products in the toroidal magnetic field in the range of laboratory polar angles from $8^\circ$
to $140^\circ$. A set of 342 time-of-flight scintillators (TOF)~\cite{tof}  was used to record the flight times of the charged
particles. Start Counter (ST) scintillators~\cite{stt} surrounded the target cell and were used to determine the event start
time. The trigger required a hit in the photon tagger in coincidence with ST and TOF hits in at least two of the six sectors of
CLAS. During the g11a run period, the total number of triggers collected was $\sim2\times10^{10}$, giving an integrated
luminosity of 70~pb$^{-1}$.

\subsection{Event selection}
\label{evsel}

We required the detection of at least two charged particles in CLAS. The event sample consisted of four mutually
exclusive topologies,  one with all three final state hadrons detected and three others in which one out of the three final
state hadrons was missing. For these events the momentum of the missing particle was reconstructed from energy-momentum
conservation. The momenta of the reconstructed charged particles were corrected for energy loss in the target materials
\cite{eloss}.  The tagged-photon energies were also corrected taking into account all known tagger focal plane mechanical
deformations~\cite{St07}.

A kinematic fit was used for the event selection to isolate the $\gamma p \to \pi^+ \pi^- p$ reaction~\cite{kinfit}. The events 
passing the kinematic fit with confidence level (CL) above 0.1 were accepted. The pull distributions of the measured 
kinematic quantities were fit by Gaussians centered at $0.00 \pm 0.05$ with $\sigma = 1.0 \pm 0.1$.

Some events passed the CL cut with one or more tracks assigned the wrong particle identity. To further clean up the event 
sample, we employed a timing cut $|T_{tag} - T_{stt}| < 1.5$~ns, where $T_{tag}$ is the vertex time of the incident photon 
measured by the tagger and $T_{stt}$ is the vertex time of the final state particle measured by the ST. The kinematic fit 
probed all matched photons, selecting the hit with the maximum CL value. The photon energy measured by the tagger 
was compared with the total energy computed from the four-momenta of the final state particles.  This energy difference 
was found to be within $\Delta E/E \approx 0.5\%$, confirming the accuracy of the detector and photon beam
calibrations and the purity of the final event sample.

The CLAS detector contained insensitive regions for particle detection. These insensitive regions were at the locations of
the torus coils, as well as at very forward ($\theta < 4^\circ$) and very backward angles ($\theta > 140^\circ$) in the lab
frame~\cite{Me03}. The final state particles were selected to be within the ``fiducial'' regions with reliable particle
detection efficiency, away from the insensitive regions. In addition, the kinematic regions where the particle detection
efficiency was less than 5\% were excluded. Overall, $\approx$400 million $\pi^+\pi^-p$ events were selected for the
evaluation of the integrated and differential cross sections exceeding by a factor of $\sim$50 the statistics previously
collected in this channel~\cite{saphir}. An uncertainty of 3\% for the event selection was determined from the mismatch
between the fraction of selected $\pi^+\pi^-p$ events in the kinematic fits of the Monte Carlo (MC) sample and the
measured data.

\subsection{Cross section evaluation}
\label{secexp} 

Studies of the $\pi^+\pi^-p$ photoproduction reaction with an unpolarized beam off an unpolarized proton target at a given
center of mass (CM) energy $W$ can be fully described by a 5-fold differential cross section. This cross section has a
uniform distribution over the azimuthal CM angles for all final state hadrons. Integrating over the azimuthal CM angle allows
the 5-fold differential cross section to be expressed as a 4-fold differential cross section. 

\begin{figure}[h]
\begin{center} 
\includegraphics[width=0.49\textwidth]{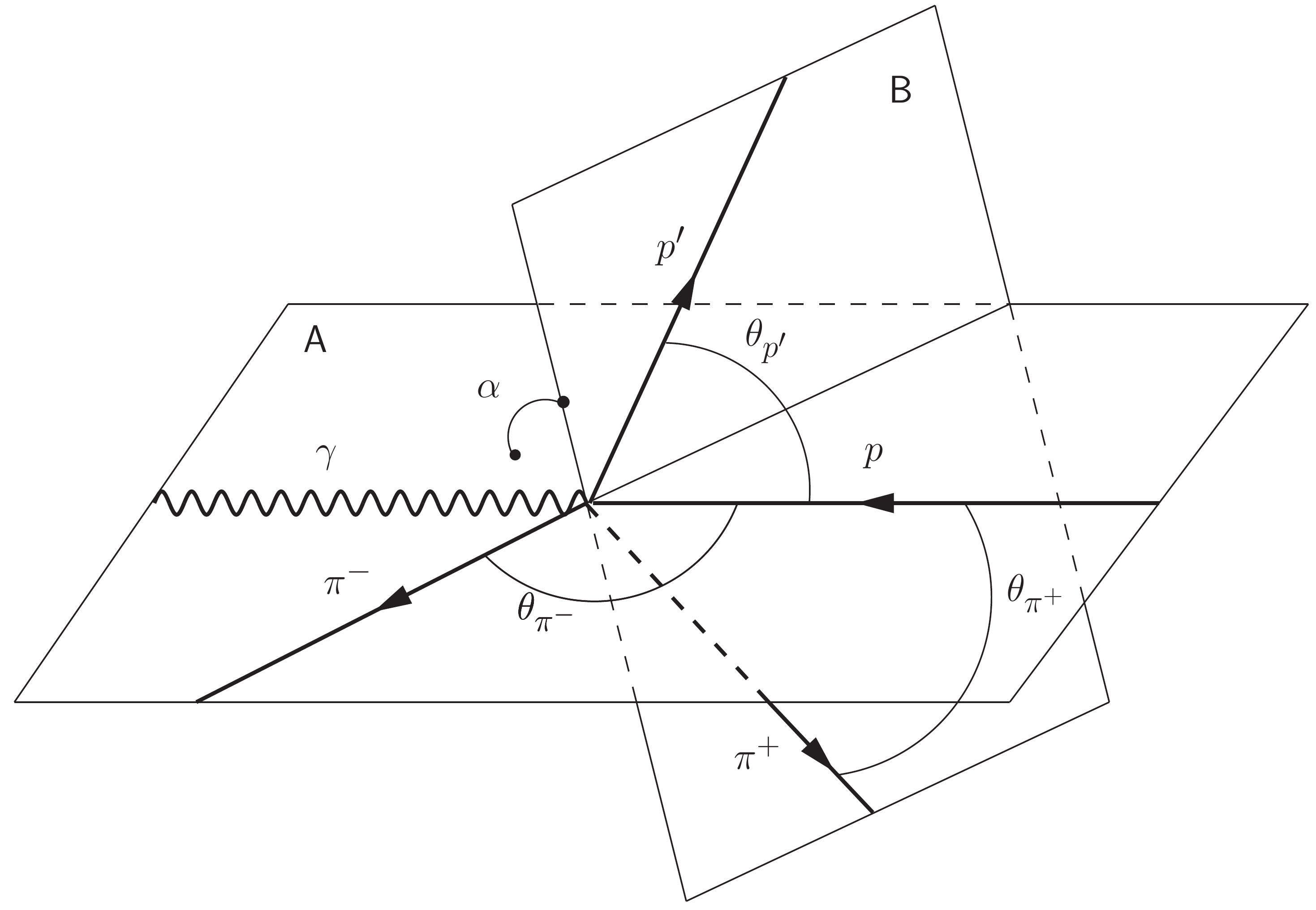} 
\caption{Angular kinematic variables for the reaction $\gamma p \to \pi^+ \pi^- p'$  in the CM frame. The variable set
with $i$=$\pi^-$, $j$=$\pi^+$, and $k$=$p'$ includes the angular variables for $\theta_{\pi^-}$, the polar angle of the
$\pi^-$, and $\alpha_{[\pi^-p][\pi^+p']}$, which is the angle between the planes $A$ and $B$, where plane $A$ ($[\pi^-p]$)
is defined by the 3-momenta of the $\pi^-$ and the initial state proton and plane $B$ ($[\pi^+p']$) is defined by the
3-momenta of the $\pi^+$ and the final state proton $p'$. The polar angle $\theta_{p'}$  is relevant for the set with
$i$=$p'$, $j$=$\pi^+$, and $k$=$\pi^-$, while the polar angle $\theta_{\pi^+}$ belongs to the set with $i$=$\pi^+$,
$j$=$p'$, and $k$=$\pi^-$.} 
\label{kin}
\end{center} 
\end{figure}

The cross sections were defined using three sets of four kinematic variables. These included the permutations of the two
invariant masses derived from pairing two of the three final state hadrons $M_{ij}$ and $M_{jk}$, where $i$, $j$, and $k$
represent the final state particles $\pi^+$, $\pi^-$, and $p'$. The definitions for the final state CM angular variables are
given in Fig.~\ref{kin}. There are two relevant CM angles in each set of variables, 1) $\theta_i$ for one of the final state
hadrons $i$ and 2) $\alpha_{[ip][jk]}$ between the two hadronic planes defined by the three-momenta of the initial state
proton $p$ and the final state hadron $i$, and the three-momenta of the remaining final state hadron pair $jk$. The reaction
kinematics are described in detail in Refs.~\cite{Mo16a,Fe09}. 

The selected $\pi^+\pi^-p$ events were sorted into 16 25-MeV-wide bins in $W$ in the range from 1.6~GeV to 2.0~GeV.
Each $W$ bin contained 16 bins in the invariant masses of the two final state hadron pairs, and 14 bins in the angles
$\theta_i$ and $\alpha_{[ip][jk]}$. The 4-fold differential cross sections were evaluated from the $\pi^+\pi^-p$ event yields
collected in the 4-dimensional (4-D) bins, normalizing by the detection efficiency in each bin and the overall beam-target
luminosity. After integration of the 4-fold differential cross sections over the three different sets of three variables, nine
1-fold differential cross sections were determined for 1.6~GeV~$< W <$~2.0~GeV. These 1-fold differential cross sections
include:

\begin{enumerate}[a)]
\item invariant mass distributions: \newline 
$\frac{d\sigma}{dM_{\pi^+p'}}$, $\frac{d\sigma}{dM_{\pi^+\pi^-}}$, $\frac{d\sigma}{dM_{\pi^-p'}}$;
\item angular distributions over $\theta$: \newline 
$\frac{d\sigma}{d(-\cos \theta_{\pi^-})}$, $\frac{d\sigma}{d(-\cos \theta_{\pi^+})}$,
$\frac{d\sigma}{d(-\cos \theta_{p'})}$;
\item angular distributions over $\alpha$: \newline 
$\frac{d\sigma}{d\alpha_{[\pi^-p][\pi^+p']}}$, $\frac{d\sigma}{d\alpha_{[\pi^+p][\pi^-p']}}$,
$\frac{d\sigma}{d\alpha_{[p'p][\pi^+\pi^-]}}$.
\end{enumerate}

Each of the nine 1-fold differential cross sections, while generated by a common 4-fold differential cross section, offers
complementary information. None of them can be computed from the others. Data on all nine 1-fold differential cross
sections are essential to gain insight into the resonant contributions of the $\pi^+\pi^-p$ reaction.

\begin{figure}[t]
\begin{center} 
\includegraphics[width=0.40\textwidth]{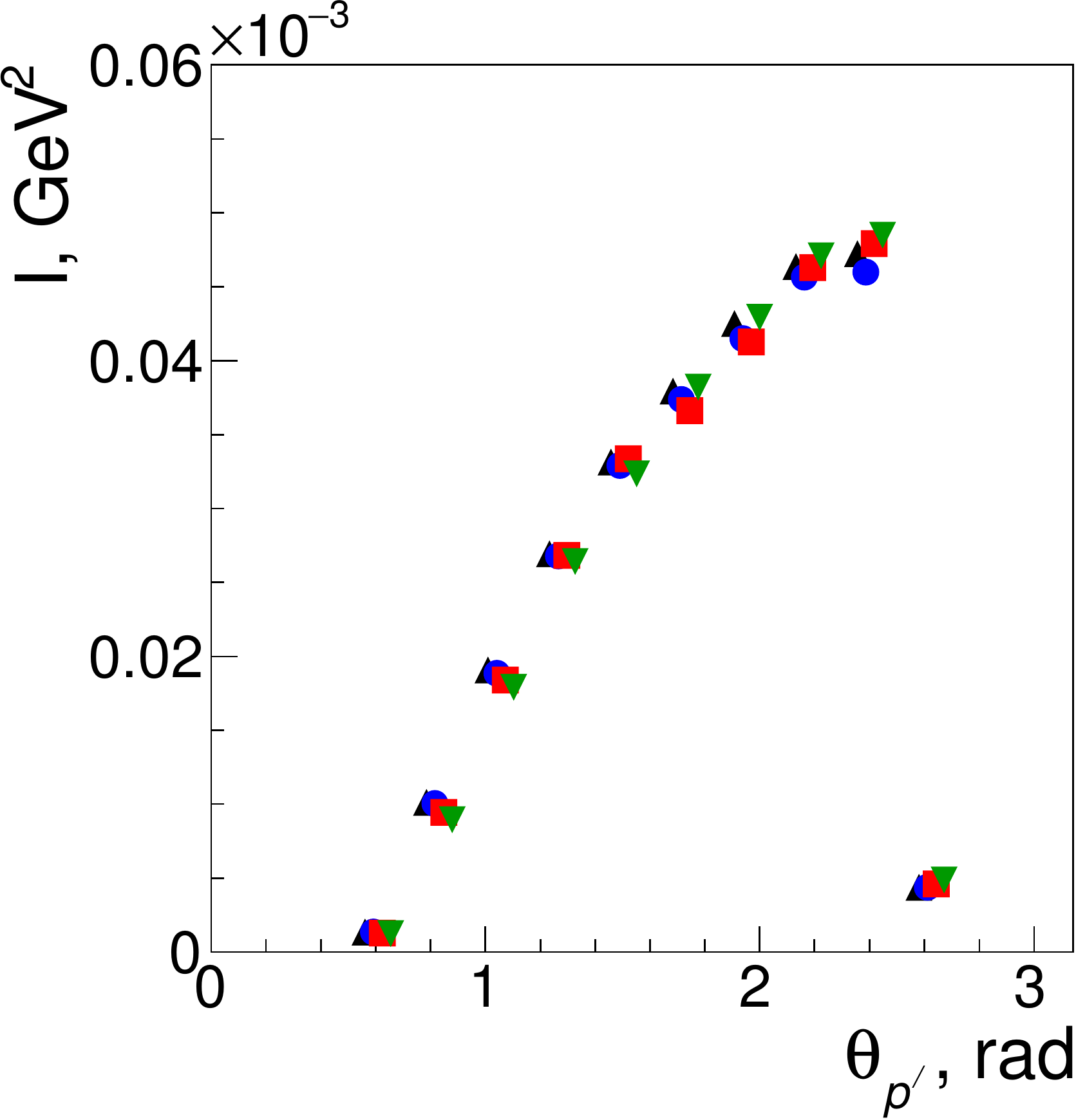}
\caption{(Color Online) Representative integrals $I$ over the variables $M_{\pi^- p'}$, $M_{\pi^+\pi^-}$, and 
$\alpha_{[p'p][\pi^+\pi^-]}$ as a function of $\theta_{p'}$ at $W$ from 1.70~GeV to 1.73 GeV defined from the
$\pi^+\pi^-p$ normalized yields in the 4-D cells.  The integrals contain only the 4-D cells where the events from
all four topologies were available. Their values for the four different topologies: all final state hadrons detected
(black triangles) and with the reconstructed momenta for the $p'$ (red squares), $\pi^-$ (blue circles),  and
$\pi^+$ (green upside down triangles). The integrals over the two invariant masses have dimensions of GeV$^2$.}
\label{tops} 
\end{center} 
\end{figure}

Parity conservation mandates that the 4-fold differential cross sections are equal at the angles $\alpha$ and
$2\pi - \alpha$. In the computation of the 1-fold differential cross sections, we have set the measured 4-fold
differential cross sections at the angles $\alpha$ and $2\pi - \alpha$ equal to their average value. This procedure
alters the 1-fold differential cross sections well within the uncertainties of
the detector efficiency. 

The detector efficiency was computed using a detailed GEANT simulation of the CLAS detector called GSIM~\cite{gsim} 
and an event generator based on the older JM05 reaction model~\cite{Ri00,Mo05}. The kinematical grid for the
reconstructed $\pi^+\pi^- p$ Monte Carlo events coincided with that described above for the measured data events.
This grid contained 802,816 5-D cells: 16 ($W$ bins) $\times$ 16 (invariant mass bins of the first final state hadron pair) 
$\times$ 16 (invariant mass bins of the second final state hadron pair) $\times$ 14 (final state hadron $\theta$ angle bins) 
$\times$ 14 (final state hadron $\alpha$ angle bins). About half of the cells resided outside of the boundary of the reaction
phase space, and such cells were removed from the analysis.  The small size of the cells allowed us to evaluate the detection
efficiency reliably even for approximate modeling of the event distributions within the JM05 model version incorporated into
the event generator. Uncertainties related to the mismatch between the actual CLAS efficiency and that determined from
the simulation were studied as discussed in Ref.~\cite{willthesis} by comparing the normalized yields of $\omega$
electroproduction events in the six sectors of CLAS. For experiments with unpolarized beam and target, all cross sections
should be uniform over the azimuthal angle. The differences between the normalized $\omega$ yields in the different CLAS
sectors was about 4\%.

The evaluation of the CLAS detection efficiency was further checked through the comparison of the integrals
of the normalized yields of the $\pi^+\pi^-p$ events for the four different final state topologies (see Section~\ref{evsel})
over the invariant masses $M_{\pi^- p'}$ and $M_{\pi^+\pi^-}$, and the angle $\alpha_{[p'p][\pi^+\pi^-]}$. The integrals were
calculated within the limited CLAS acceptance region where the 4-D cells contained the selected events of all four topologies.
The four integrals $I$ were obtained in each bin of $W$ as a function of the CM angle $\theta_{p'}$. The deviation of the
integrals from the four different topologies was about 4\%. This variation was assigned as the systematic uncertainty for the
detection efficiency (see Table~\ref{system_uncert}). A representative example for comparison between the values of the
four integrals is shown in Fig.~\ref{tops}.    

The tagged photon flux on the target within the data acquisition live time was obtained by the standard CLAS {\it gflux} 
method~\cite{flux}. The number of photons for each tagger counter was calculated independently as $N_\gamma = 
\epsilon \cdot N_{e^-}$, where $N_{e^-}$ is the number of electrons detected by a tagger counter and $\epsilon$ is the 
tagging ratio. The tagging ratio was determined by placing a total absorption counter directly in the photon beam at 
low intensity and determining the ratio of the number of beam photons and the number of electrons detected in coincidence 
in the tagger. The global normalization uncertainty derived from the run-to-run variance and the estimated normalization 
variance with the electron beam current together were found to be 3.5\%, employing the method described in 
Ref.~\cite{willthesis}.

In the determination of the fully integrated and 1-fold differential cross sections, the contributions from the insensitive
areas of CLAS were taken into account by extrapolating the 4-fold differential cross sections. As a starting point, the
evaluation of the 1-fold differential cross sections in the full acceptance was carried out in the following way. The cross
section values determined in each one-dimensional (1-D) bin within the CLAS acceptance were multiplied by the ratio of the
total number of 4-D bins that contributed to the analyzed 1-D bin to the number of bins with non-zero efficiency (cross
section set \#1). 

\begin{figure}[h]
\begin{center}
\includegraphics[width=0.49\textwidth]{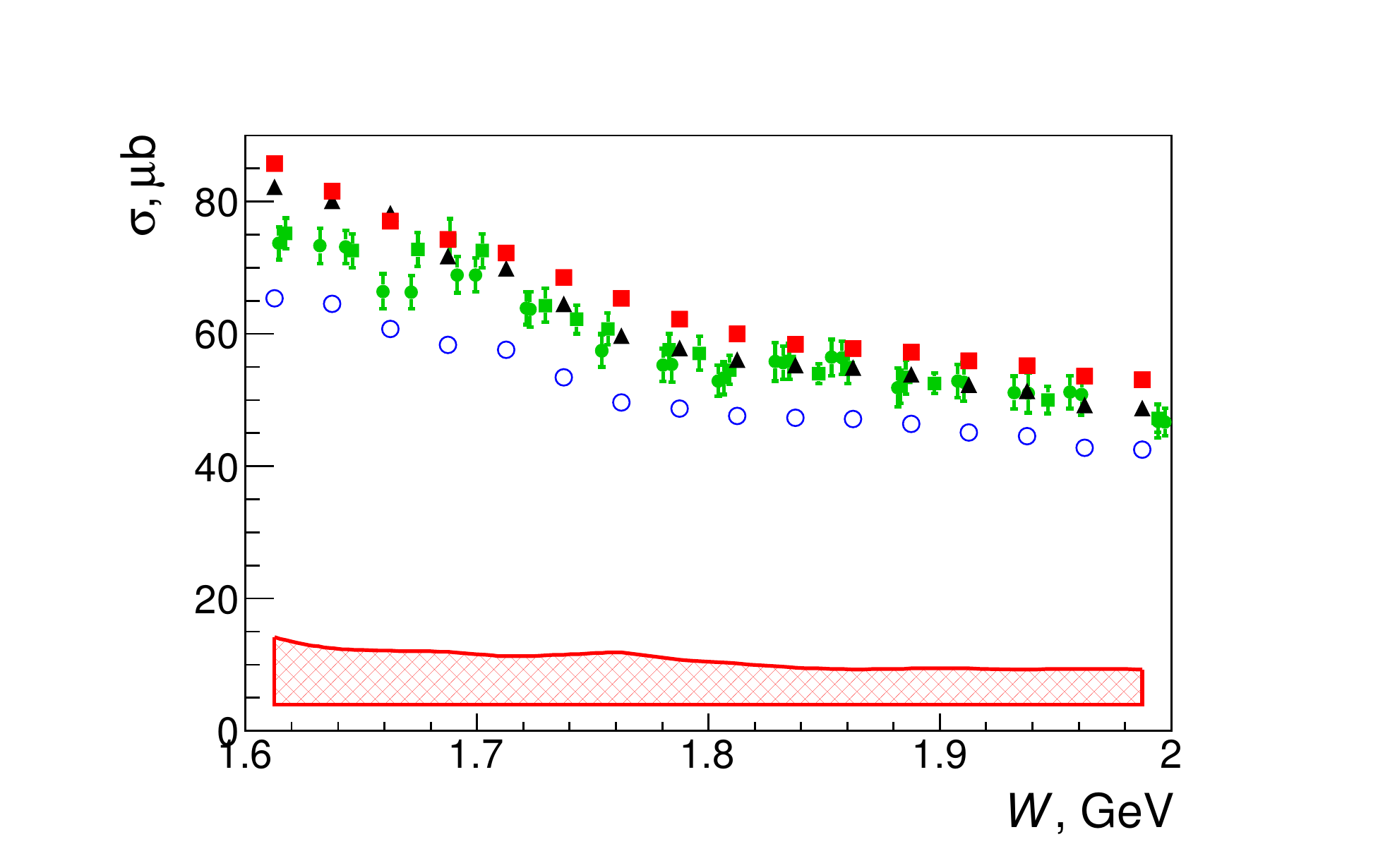} 
\caption{(Color Online) Fully integrated $\pi^+\pi^- p$ photoproduction cross sections within the CLAS acceptance (blue 
open circles) and in the full acceptance after the initial 4-fold differential cross section extrapolation into the insensitive
areas (black triangles - cross section set \#1) and after the improved extrapolation within the framework of the JM17 model
as described in Section~\ref{secexp} (red squares - cross section set \#3). The CLAS data are compared with the SAPHIR
\cite{saphir} (green squares with error bars) and the ABBHHM~\cite{abbhhm} (green circles with error bars) results. The
statistical uncertainties of our data are smaller than the marker size, while the systematic uncertainties are shown by the
hatched area at the bottom of the figure.}
\label{integ}
\end{center} 
\end{figure}

\begin{figure}
\begin{center} 
\includegraphics[width=0.36\textwidth]{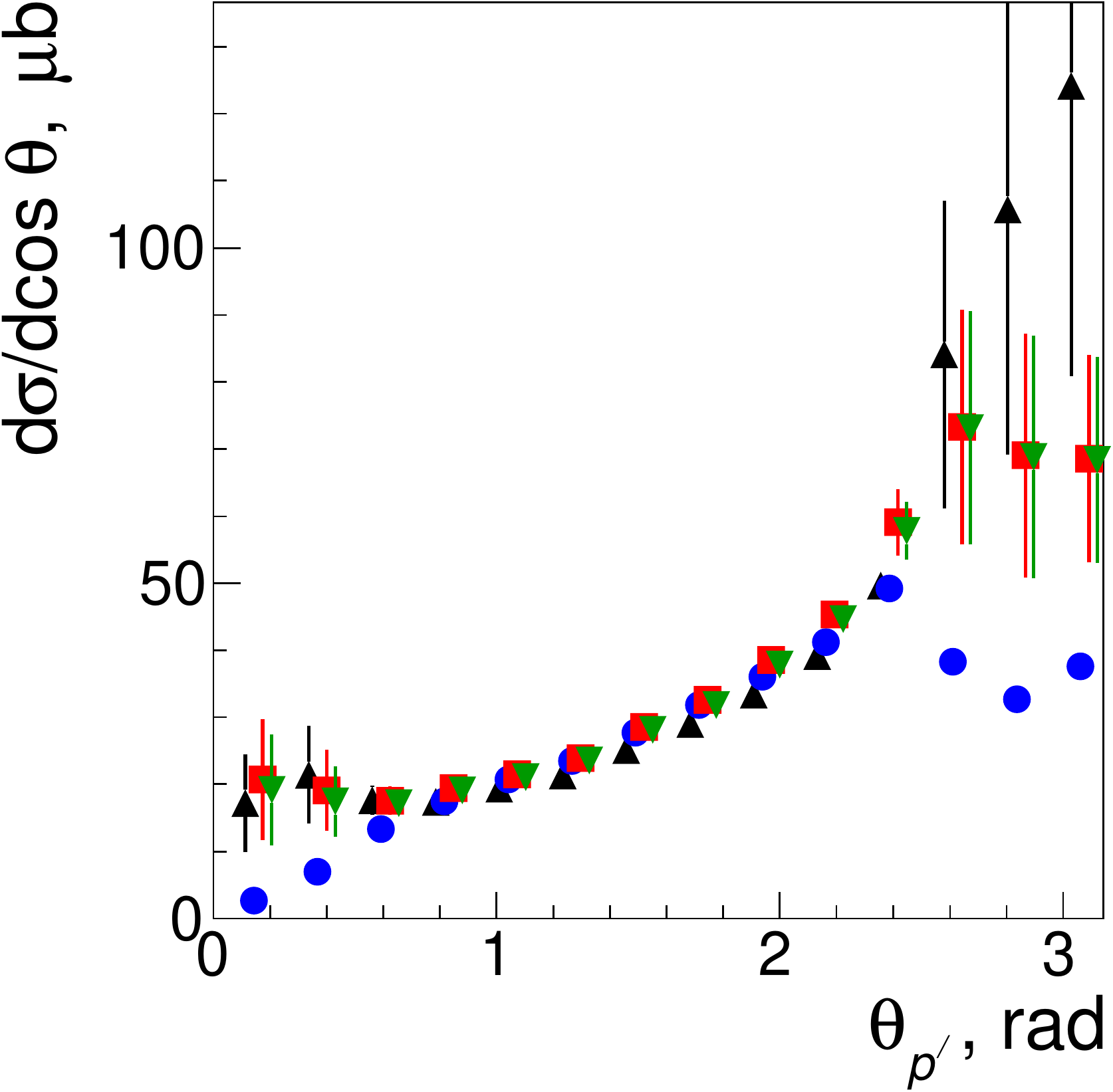}
\vspace{-0.1cm}
\caption{(Color Online) Representative $\theta_{p'}$ angular distributions at $W$ from 1.70~GeV to 1.73~GeV. Results
are obtained within the CLAS acceptance (blue circles) and in the full acceptance extrapolating the cross section into the
insensitive areas after the initial cross section extrapolation (black triangles - cross section set \#1) and with the improved
extrapolation using the JM17 model (red squares - cross section set \#3) as explained in Section~\ref{secexp}. The
results obtained by extrapolating the cross section into the insensitive areas with the initial JM17 model parameters
(cross section set \#2) are shown by the green upside down triangles. The error bars are dominated by the uncertainty
of the extrapolation procedure.}
\label{diff}
\end{center} 
\end{figure}

An improved extrapolation of the 4-fold $\pi^+\pi^-p$ differential cross sections into the insensitive areas of CLAS was
carried out within the framework of the new JM17 model described in Section~\ref{expres}. The JM17 model parameters
were fit to the data within the CLAS acceptance and the 4-fold differential cross sections in the insensitive areas were
computed from the JM17 model (cross section set \#2). Then, the JM17 model parameters were refit to reproduce the
cross sections determined in the full acceptance, obtained after filling the insensitive areas. The JM17 model with improved
parameters was then used again for the evaluation of the cross sections in the insensitive areas of CLAS, generating a new
set of differential cross sections extrapolated into the insensitive areas of CLAS (cross section set \#3). The uncertainties
caused by this cross section extrapolation were assigned as half the difference between the cross sections determined within
the full and CLAS acceptances, which amounted to 12.0\% for the integrated cross sections. This uncertainty is strongly
dependent on the CM polar angles of the final state hadrons. It was found that the two sets of nine 1-fold differential cross
sections in the full acceptance agreed within the uncertainties of the data, accounting for both the statistical and
extrapolation uncertainties.  Since the cross section of the JM17 model accounts for the amplitude constraints imposed by
parity conservation, the 4-fold differential cross sections at angles $\alpha$ and $2\pi - \alpha$ are equal within the
insensitive areas.

Figure~\ref{integ} shows the fully integrated cross section within the CLAS acceptance (blue circles). The other points
(black triangles and red squares) are the cross sections within the full acceptance after the initial and improved
cross section extrapolations into the insensitive areas of CLAS. Fig.~\ref{diff} shows a representative example of the
different cross section angular distributions from the initial cross section set \#1 and the two subsequent cross section
sets \#2 and \#3 after the improved extrapolations into the insensitive areas for the $\theta_{p'}$ CM angular distributions. 

The systematic uncertainties related to the selection of the fiducial areas were estimated by comparing the cross sections
computed with two different minimum CLAS detection efficiency cuts: 5\% (nominal) and 10\% (increased). The 4-fold
differential cross section inside the excluded areas with small detection efficiency was estimated within the extrapolation
procedure described above. The computed cross sections with the increased and nominal detection efficiency cuts differ by
about 4\% as listed in Table~\ref{system_uncert}.

The systematic uncertainties for the fully integrated $\pi^+\pi^-p$ photoproduction cross sections are summarized in
Table~\ref{system_uncert}. The largest contribution comes from the 4-fold differential cross section extrapolation into
the insensitive areas of CLAS.  

\begin{table} 
\begin{center} 
\begin{tabular}{|c|c|} \hline 
Source of uncertainty    & \begin{tabular}{@{}c@{}}
                           Contribution to fully \\ 
                           integrated $\pi^+\pi^-p$ \\
                           cross section, \%
			   \end{tabular}	   \\ \hline

Fiducial area choice   & 4.0 \\
Event selection          & 3.0 \\ 

\begin{tabular}{@{}c@{}}
Run-to-run stability and\\ 
global normalization factor
\end{tabular}            & 3.5 \\ \hline

\begin{tabular}{@{}c@{}}
Efficiency from MC\\ 
\end{tabular}            & 4.0 \\ 

\begin{tabular}{@{}c@{}}
Impact of the CLAS\\ 
insensitive areas
\end{tabular}            & 12.0 \\ \hline 

Total                    & 14.0 \\ \hline 
\end{tabular}
\caption{Summary of the systematic uncertainties for the fully integrated $\pi^+\pi^- p$ photoproduction cross sections.
The scale uncertainties and point-to-point uncertainties are listed in the second and third rows, respectively.}
\label{system_uncert}
\end{center} 
\end{table}

\section{Results and Physics Analysis} 
\label{expres}

\begin{figure*} 
\begin{center}
\includegraphics[width=0.45\textwidth]{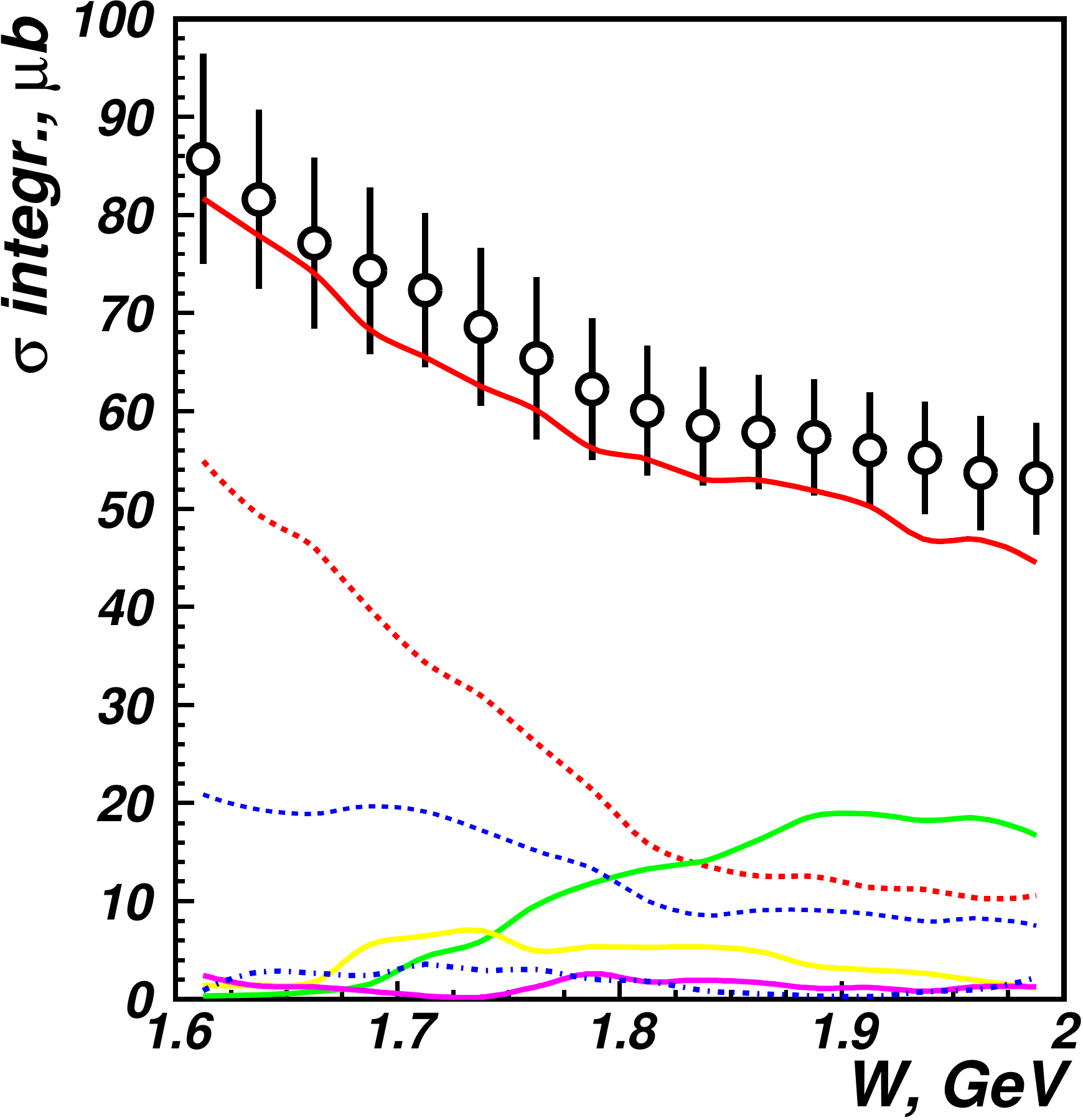}
\includegraphics[width=0.47\textwidth]{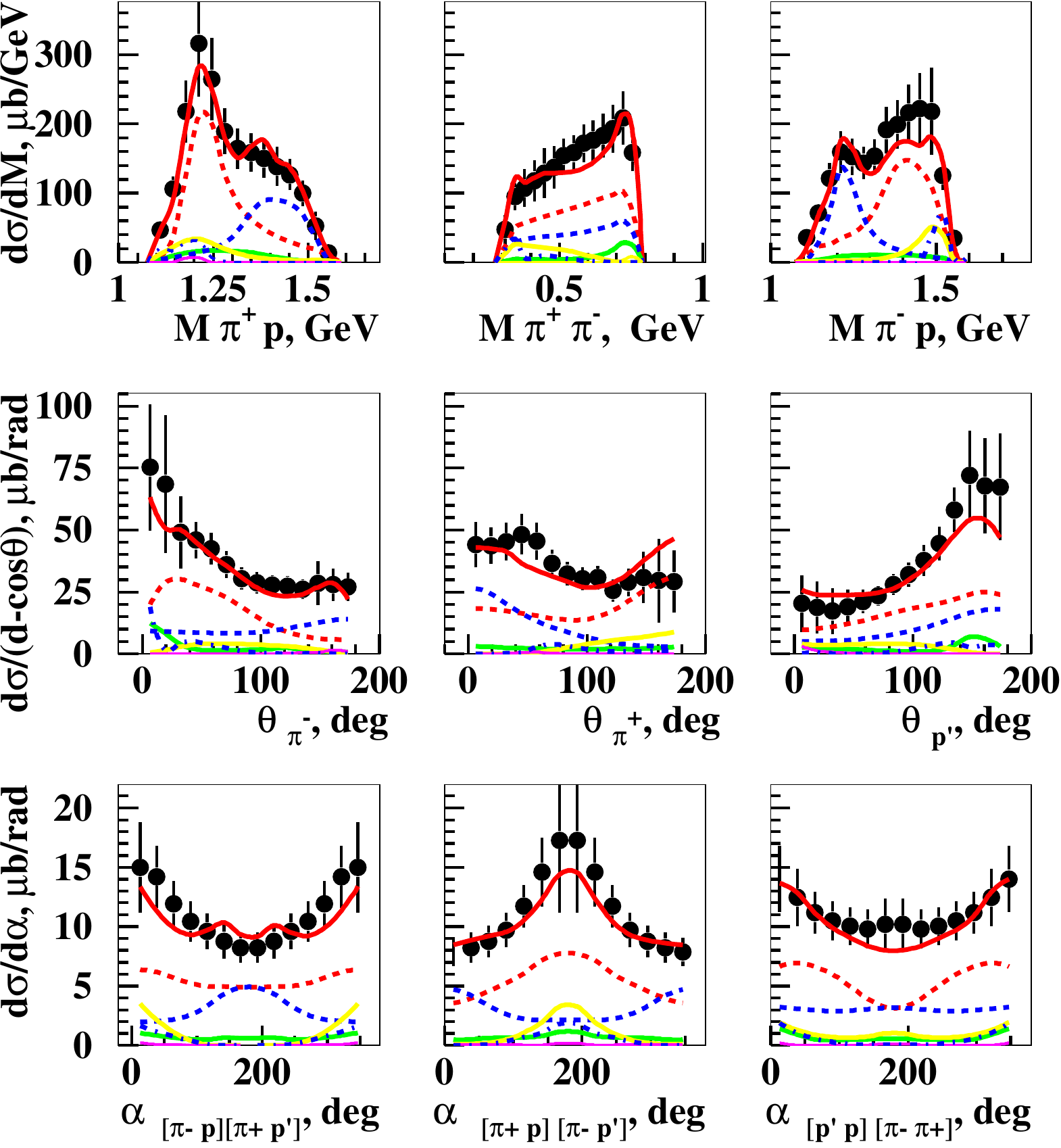}
\vspace{-0.1cm}
\caption{(Color Online) Description of the $\pi^+\pi^-p$ photoproduction cross sections and the contributions from the
relevant channels inferred from the CLAS data within the framework of the JM17 model for the fully integrated cross
sections (left) and a representative example of the nine 1-fold differential cross sections at $W$ from 1.70~GeV to
1.73~GeV (right) shown by different lines: full reaction cross sections (solid red), $\pi^-\Delta^{++}$ (dashed red),
$\rho p$ (solid green), $\pi^+\Delta^0$ (dashed blue),  $\pi^+ N(1520)3/2^-$ (yellow), 2$\pi$ direct production
(magenta), and $\pi^+ N(1685)5/2^+$ (blue dot-dashed). The error bars include the combined statistical
and point-to-point systematic uncertainties.}
\label{1diffint}
\end{center} 
\end{figure*}

\begin{figure*} 
\begin{center}
\includegraphics[width=0.45\textwidth]{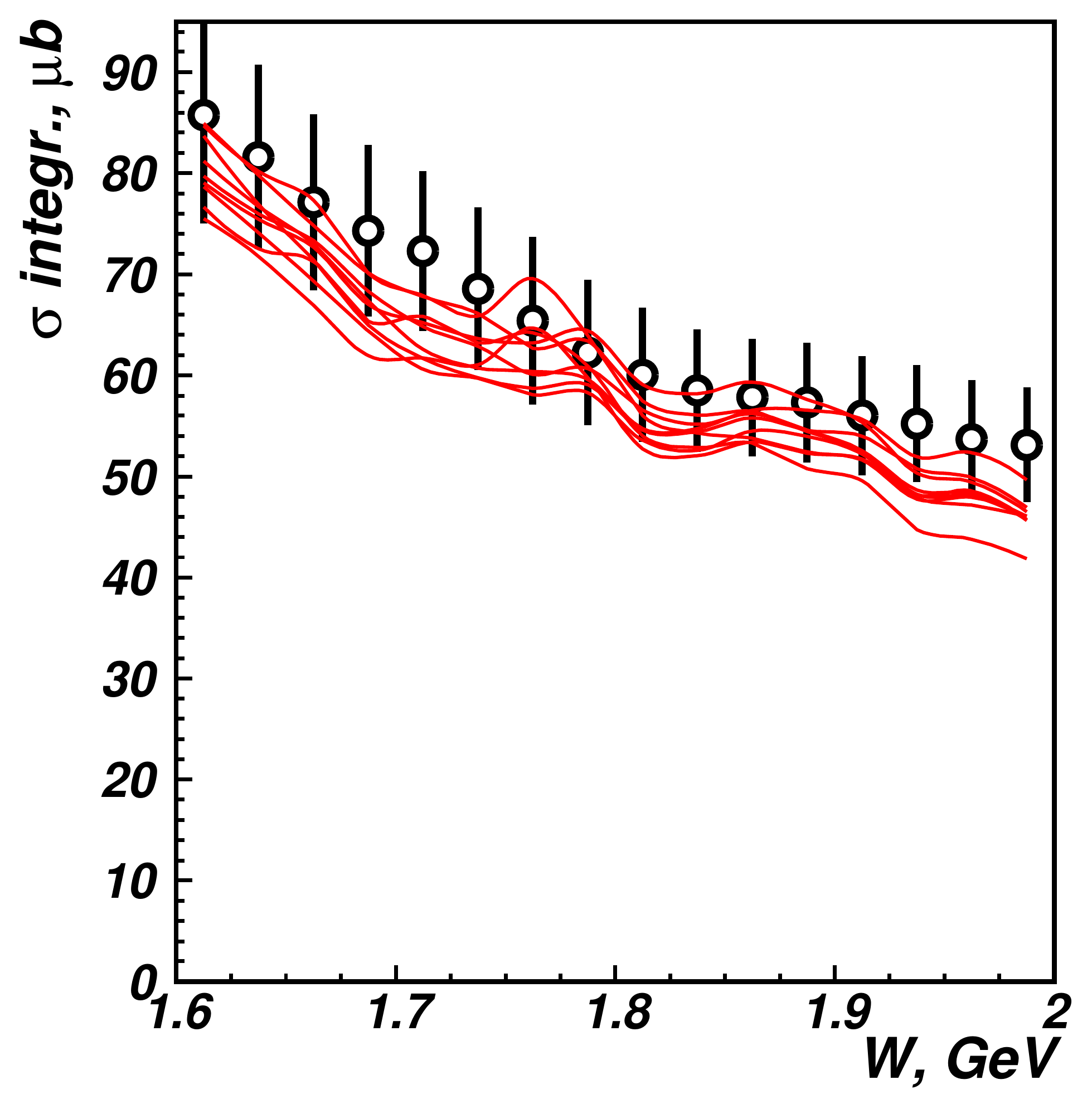}
\includegraphics[width=0.47\textwidth]{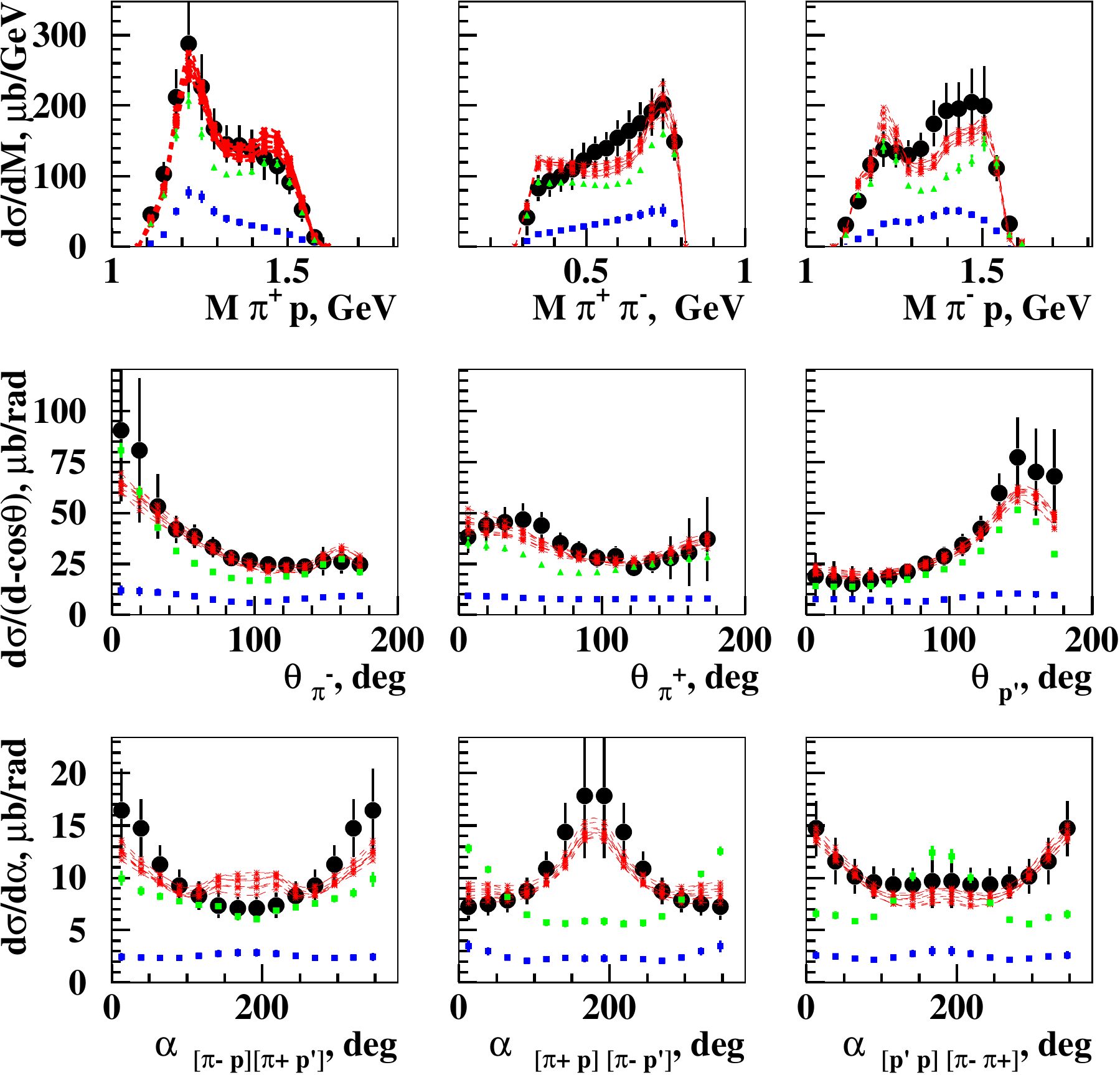}
\vspace{-0.1cm}
\caption{(Color Online) (Left) Fully integrated cross sections computed from the fits of the nine 1-fold differential cross
sections with $1.15 < \chi^2/d.p. < 1.30$ (red curves) in comparison with the measured integrated cross sections (points with
error bars). The error bars include the combined statistical and point-to-point systematic uncertainties. (Right) Representative
example of 1-fold differential cross sections (red curves) and the resonant/non-resonant contributions (blue/green bars) from
the fits with $1.15 < \chi^2/d.p. < 1.30$ of the CLAS $\pi^+\pi^-p$ photoproduction data at $W$ from 1.73~GeV to 1.75~GeV
within the framework of the JM17 model. The quoted ranges for the resonance parameters were obtained from the sets of fits
that resulted in $1.15 < \chi^2/d.p. < 1.30$ and shown by the red curves.}
\label{1diffint_resback}
\end{center} 
\end{figure*}

The fully integrated $\pi^+\pi^- p$ photoproduction cross section and representative examples of the nine 1-fold
differential cross sections are shown in Fig.~\ref{integ}, Fig.~\ref{1diffint}, and Fig.~\ref{1diffint_resback}, respectively.
We show the differential cross sections in the $W$ bins centered at 1.71~GeV and 1.74~GeV, which correspond to
the peak region of the resonance-like structure observed in the $W$ dependence of the $\pi^+\pi^-p$ electroproduction
cross sections~\cite{Ri03}. The complete set of differential cross sections from this experiment can be found in the CLAS
physics database~\cite{clasdb}. The error bars for the cross sections shown in Figs.~\ref{1diffint} and \ref{1diffint_resback}
include the uncertainties related to the extrapolation of the 4-fold differential cross sections into the insensitive areas of
CLAS. The fully integrated cross sections from CLAS are consistent with the existing results within the systematic
uncertainties~\cite{saphir,abbhhm}. However, our fully integrated cross sections in the full acceptance are slightly above the
existing results likely due to the different approaches used for the cross section extrapolations into the insensitive areas. We
consider estimates of the 5-fold differential cross sections in the insensitive areas from the JM17 model, outlined below, as
reliable, since the nine 1-fold differential cross sections are well described within the acceptance as shown in Figs.~\ref{1diffint}
and \ref{1diffint_resback}.

Data on the angular distributions over the three $\alpha$ angles described in Section~\ref{secexp} have become
available for the first time. Also differential cross sections over the final state hadron CM $\theta_{i}$ angles
($i$=$\pi^+$, $\pi^-$, $p$) offer information on the reaction dynamics different from the distributions over the
Mandelstam $t$ variable included in Ref.~\cite{saphir}. The first results on the nine 1-fold differential cross sections
make it possible to isolate the resonant contributions to the $\pi^+\pi^-p'$ reaction and to determine the resonance
photocouplings from this channel.

\begin{table}
\begin{center}
\begin{tabular}{|c|c|c|c|} \hline
  Resonance    & Mass,      & Total  &  Branching  \\
               & GeV        & width, &  fraction to \\
                     &      & GeV    & $\pi\pi N$, \%  \\ \hline
$N(1440)1/2^+$       & 1.45 &  0.35  &   37   \\
$N(1520)3/2^-$       & 1.52 &  0.13  &   41   \\
$N(1535)1/2^-$       & 1.53 &  0.15  &    4   \\
$\Delta(1620)1/2^-$  & 1.63 &  0.15  &   93   \\
$N(1650)1/2^-$       & 1.65 &  0.14  &    7   \\
$N(1680)5/2^+$       & 1.69 &  0.14  &   35   \\
$\Delta(1700)3/2^-$  & 1.70 &  0.30  &   86   \\
$N(1720)3/2^+$       & 1.74 &  0.12  &   85   \\
$N'(1720)3/2^+$      & 1.72 &  0.12  &   68   \\
$\Delta(1905)5/2^+$  & 1.88 &  0.33  &   87   \\
$\Delta(1910)1/2^+$  & 1.89 &  0.28  &   12   \\
$\Delta(1950)7/2^+$  & 1.93 &  0.29  &   61   \\
$N(2190)7/2^-$       & 2.19 &  0.50  &   40   \\ \hline
\end{tabular}
\caption{Starting values for the hadronic decays parameters of the excited nucleon states incorporated into the JM17
model version for the description of the $\pi^+\pi^-p$ photoproduction data.}
\label{reslist} 
\end{center}
\end{table}

The photocouplings of the nucleon resonances in the mass range from 1.6~GeV to 2.0~GeV were determined from a fit to all
nine 1-fold differential cross sections from $\pi^+\pi^-p$ photoproduction. First, we established the relevant
mechanisms contributing to this exclusive channel from their manifestations in the observables. The observable description
was performed starting from the JM16 model~\cite{Mo12,Mo16a} updated to describe the $\pi^+\pi^-p$ photoproduction
data (JM17 model). The previous JM model versions, described in Refs.~\cite{Mo12,Mo16a,Mo09}, were successfully used
for the extraction of the nucleon resonance electrocouplings from the CLAS $\pi^+\pi^- p$ electroproduction data
\cite{Mo16}. The JM17 model incorporates all mechanisms that contribute
to $\pi^+\pi^-p$ electroproduction in the resonance region with manifestations seen in the measured differential
photoproduction cross sections. These consist of the $\pi^-\Delta^{++}$, $\pi^+\Delta^0$, $\rho^0 p$, $\pi^+N(1520)3/2^-$,
and $\pi^+N(1685)5/2^+$ meson-baryon channels, as well as the direct production of the $\pi^+\pi^-p$ final state without
formation of intermediate unstable hadrons. The modeling of these processes was described in
Refs.~\cite{Mo12,Mo09,Mo16,Ri00,Mo05}.

The differences in the kinematic dependence of the $\alpha_{[\pi^-p][\pi^+p']}$ angular distributions for $\pi^+\pi^-p$ photo-
and electroproduction were accounted for in the phenomenological parameterization of the direct 2$\pi$ production
mechanisms of Ref.~\cite{Mo09}. The $\pi^+\pi^-p$ photoproduction data at $W > 1.8$~GeV require implementation of the
$\sigma p$ meson-baryon channel, which was parameterized by a 3-body contact term and an exponential propagator for the
intermediate $\sigma$ meson. The magnitudes of the parameterized $\sigma p$ photoproduction amplitudes were fit to the
data in each bin of $W$ independently. The contributions from all well established $N^*$ states with masses $< 2.0$~GeV
with observed decays to the $\pi\pi N$ final states (listed in Table~\ref{reslist}) were included into the $\pi\Delta$ and
$\rho p$ meson-baryon channels of JM17. The resonant amplitudes were described in a unitarized Breit-Wigner ansatz
\cite{Mo12} that accounted for restrictions imposed by a general unitarity condition to the resonant contributions
\cite{Ait72}.

The initial values for the $\pi\Delta$ and $\rho p$ decay widths were taken from analyses of the previous CLAS
$\pi^+\pi^-p$ electroproduction data~\cite{Mo12,Mo16a} for the $N(1440)1/2^+$, $N(1520)3/2^-$, and
$\Delta(1620)1/2^-$ resonances. For other $N^*$ states in the mass range up to 2.0~GeV, we used the results of
Ref.~\cite{Rpp18} for the total decay width and from Ref.~\cite{Man92} for the $\pi\Delta$ and $\rho p$ decay widths.
The parameters for the $N(2190)7/2^-$ resonance were taken from Ref.~\cite{Rpp18}. The initial resonance photocouplings
were taken from Refs.~\cite{Rpp18,Mo16b,Dug09}. Independent fits of the $\pi^+\pi^-p$ photo- and electroproduction
\cite{Ri03} cross sections assuming the contributions from the known resonances only, result in a factor of four difference
of the branching fractions for the decays of the conventional $N(1720)3/2^+$ resonance to the $\rho N$ final state. Since
resonance decay widths should be $Q^2$ independent, it is impossible to describe both the $\pi^+\pi^-p$ photo- and
electroproduction cross sections when only contributions from conventional resonances are taken into account. By implementing
a new $N'(1720)3/2^+$ state with the mass, photo- and hadronic couplings starting from the values in Ref.~\cite{Mo16b}, a
successful description of all $\pi^+\pi^-p$ differential cross sections for photo- and electroproduction  was achieved with
$Q^2$ independent hadronic decays for the included resonances located at $W \approx 1.7$~GeV, thus validating the 
contribution from the $N'(1720)3/2^+$ state~\cite{Mo16b}. 

Before extraction of the nucleon resonance photocouplings, we validated the mechanisms incorporated into the JM17 model
(described above) by confronting the model expectations and the measured cross sections. We consider the successful
description of the nine 1-fold differential cross sections as strong evidence for the proper accounting of all essential
reaction contributions. We computed the nine 1-fold cross sections, as well as the contributions from all mechanisms
incorporated into the JM17 model, with the model parameters adjusted to reproduce the data. A similar approach was used
successfully for the extraction of the $\gamma_{v}pN^*$ electrocouplings from the $\pi^+\pi^-p$ electroproduction data
\cite{Mo12, Mo16a,Fe09} included in the PDG~\cite{Rpp18}. The JM17 model reproduces well the $\pi^+\pi^- p$ differential
cross sections for $W < 2.0$~GeV (see Figs.~\ref{1diffint} and \ref{1diffint_resback}), with a $\chi^2$ per data point
($\chi^2/d.p$) in individual $W$ bins less than 1.4.

\begin{table*}
\begin {center}
\begin{tabular}{ c  
                 D{,}{\pm}{-1}  
		 D{,}{\hspace{3pt} \text{--} \hspace{3pt}}{-1} 
                 D{,}{\pm}{-1}   
		 D{,}{\pm}{-1} 
		 D{,}{\hspace{3pt} \text{--} \hspace{3pt}}{-1}   
		 D{,}{\pm}{-1}}
\toprule
{}
& \multicolumn{1}{l}{$A_{1/2}\times 10^3$}
& \multicolumn{1}{l}{$A_{1/2}\times 10^3$}
& \multicolumn{1}{l}{$A_{1/2}\times 10^3$}
& \multicolumn{1}{l}{$A_{3/2}\times 10^3$}
& \multicolumn{1}{l}{$A_{3/2}\times 10^3$}
& \multicolumn{1}{l}{$A_{3/2}\times 10^3$}
\\
{Resonances}
& \multicolumn{1}{l}{from $\pi^+\pi^-p$}
& \multicolumn{1}{l}{PDG ranges}
& \multicolumn{1}{l}{multichannel}
& \multicolumn{1}{l}{from $\pi^+\pi^-p$}
& \multicolumn{1}{l}{PDG ranges}
& \multicolumn{1}{l}{multichannel}
\\
{}
& \multicolumn{1}{l}{GeV$^{-1/2}$}
& \multicolumn{1}{l}{GeV$^{-1/2}$}
& \multicolumn{1}{l}{analysis~\cite{So17}}
& \multicolumn{1}{l}{GeV$^{-1/2}$}
& \multicolumn{1}{l}{GeV$^{-1/2}$}
& \multicolumn{1}{l}{analysis~\cite{So17}}
\\
& \multicolumn{1}{l}{}
& \multicolumn{1}{l}{}
& \multicolumn{1}{l}{GeV$^{-1/2}$}
& \multicolumn{1}{l}{}
& \multicolumn{1}{l}{}
& \multicolumn{1}{l}{GeV$^{-1/2}$}
\\
\midrule
$\Delta(1620)1/2^-$    & 29.0,6.2    &	30 , 60     &   55,7      & 	        &  	        &	     \\
$N(1650)1/2^-$         & 60.5,7.7    &	35 , 55     &   32,6      & 	        &  	        &	     \\
$N(1680)5/2^+$         & -27.8,3.6   &  -18 ,{-5}   &  -15,2      & 128,11      &  130 , 140    &  136 ,  5  \\
$N(1720)3/2^+$         & 80.9,11.5   &  80 , 120    &   115,45    & -34.0,7.6   &  -48 , 135    &  135 , 40  \\
$\Delta(1700)3/2^-$    & 87.2,18.9   &  100 , 160   &   165,20    & 87.2,16.4   &   90 , 170    &  170 , 25  \\
$\Delta(1905)5/2^+$    & 19.0,7.6    &  17 , 27     &   25,5      & -43.2,17.3  &  -55 , {-35}  &  -50  , 5  \\
$\Delta(1950)7/2^+$    & -69.8,14.1  &  -75 , {-65} &   -67,5     & -118.1,19.3 &  -100 , {-80} &  -94 , 4   \\ 
\bottomrule	        	      
\end{tabular}
\caption{Resonance photocouplings determined from analysis of the $\pi^+\pi^-p$ photoproduction data from this work in 
comparison with the previous results from the PDG average \cite{Rpp18} and from multichannel analysis \cite{So17}.}
\label{nstpar}
\end {center}
\end{table*}

As shown in Fig.~\ref{1diffint}, the individual contributing mechanisms have distinctive differences in the shapes in all nine 
1-fold differential cross sections. The details of the shapes of these contributions are determined by the underlying reaction 
dynamics.  Therefore, the successful reproduction of the measured cross sections within the JM17 model provides confidence
that this model incorporates all essential contributing mechanisms and offers a reasonable description of them. Furthermore,
this agreement provides strong confidence that this model can be used for the extraction of the resonance parameters.

The resonance photocouplings and the $\pi \Delta$ and $\rho p$ decay widths were determined from the data
fit, where they were varied independently together with the parameters of the non-resonant amplitudes described in
Refs.~\cite{Mo12,Mo16a} and the magnitudes of the $\sigma p$ photoproduction amplitudes. These parameters were
sampled around their initial values, employing unrestricted normal distributions with a width ($\sigma$) of magnitude 30\%
of their initial values. Under this variation, the computed 1-fold differential cross sections overlap the measured cross
sections within the combined statistical uncertainties and point-to-point systematic uncertainties. In this way, we scanned the
full space of the JM17 model resonant and non-resonant parameters that provided comparable computed cross sections
with the data. For each trial set of the fit parameters, we computed the nine 1-fold differential $\pi^+\pi^-p$ cross sections
and estimated the $\chi^2/d.p.$ values in point-by-point comparisons. The resonance photocouplings and the $\pi \Delta$
and $\rho p$ decay widths were recorded from the fits over the entire $W$ range from 1.6~GeV to 2.0~GeV that
resulted in $1.15 < \chi^2/d.p. < 1.3$. This $\chi^2/d.p.$ range amounts to requiring that the computed cross sections from
the fits be within the data uncertainties.

The robustness of the fit is demonstrated in Fig.~\ref{1diffint_resback} where the selected computed differential cross
sections together with the resonant/non-resonant contributions are shown for $W$=1.74~GeV as a typical example. From the
selected fits, the uncertainties of the resonant contributions are comparable with those for the experimental data,
suggesting unambiguous access to these contributions in the differential cross sections. The resonance photocouplings
were determined from the resonant contributions by employing a unitarized Breit-Wigner ansatz~\cite{Mo12}. The
differences of the resonant and non-resonant contributions (see Fig.~\ref{1diffint_resback}) in the nine 1-fold differential
cross sections, in particular in the CM angular distributions, allows clean resonance photocoupling extraction even in bins where
the resonance contribution is smaller than the non-resonant contribution. 

The resonance parameters determined from the fits that fell within our defined $\chi^2/d.p.$ range were averaged and
their mean values were taken as the extracted resonance parameters. The dispersion in these parameters was taken as the
associated systematic uncertainty.

The resonance photocouplings extracted from this work are listed in Table~\ref{nstpar} and compared with the resonance
photocoupling ranges and the results of the multichannel analysis included in the PDG~\cite{Rpp18}. Our results were
obtained with the resonance masses, total widths, and branching fractions to the $\pi\pi N$ final states ($\beta_{\pi\pi N}$)
listed in Table~\ref{hadrpar}. The ranges of the branching fractions were computed from the ranges of the resonance total
($\Gamma_{tot}$) and partial decay widths to the $\pi\pi N$ final states ($\Gamma_{\pi\pi N}$) obtained in the data fit. The
$\Gamma_{tot}$ ranges listed in Table~\ref{hadrpar} were computed as $\Gamma_{\pi\pi N}/\beta_{\pi\pi N}$ with the mean
$\Gamma_{\pi\pi N}$ values from the data fit and the $\beta_{\pi\pi N}$ ranges from the last column of Table~\ref{hadrpar}.
For the resonances with masses below 1.8~GeV, we employed additional constraints on the total and the $\pi\pi N$ partial
decay widths from the successful combined fit of the $\pi^+\pi^-p$ photo- and electroproduction data~\cite{Mo16b,Ri03,Mo14}
with $Q^2$-independent resonance masses and decay widths. This powerful constraint considerably improved knowledge on the
$N^*$  total and $\pi\pi N$ partial decay widths, as can be seen in Table~\ref{hadrpar} from the comparison of the decay
parameter uncertainties for resonances below 1.8~GeV to those with masses above 1.8~GeV.

There is good agreement in the magnitude and sign of the photocouplings between our results and the photocoupling
ranges in the PDG listings. On the other hand, for several resonances in Table~\ref{nstpar}, the photocouplings determined
from the multichannel analysis of Ref.~\cite{So17} are different from ours. Implementation of our $\pi^+\pi^-p$
photoproduction data into the multichannel analyses will allow for examination of these differences and to improve our
knowledge on the photocouplings and hadronic decay parameters of the resonances listed in Tables~\ref{nstpar} and
\ref{hadrpar}.

\begin{table*}
\begin{center}
\begin{tabular}{|c|c|c|c|} \hline
  Resonance    & Mass,      & Total  &  Branching  \\
               & GeV        & width, &  fraction to \\
                     &      & GeV    & $\pi\pi N$, \%  \\ \hline

$\Delta(1620)1/2^-$  & 1.635 $\pm$ 0.008 &  0.144 $\pm$ 0.016  &   81-100   \\
$N(1650)1/2^-$       & 1.657 $\pm$ 0.006 &  0.154 $\pm$ 0.028  &   11-14   \\
$N(1680)5/2^+$       & 1.686 $\pm$ 0.005 &  0.118 $\pm$ 0.020  &   20-28   \\
$N(1720)3/2^+$       & 1.745 $\pm$ 0.006 &  0.116 $\pm$ 0.027  &   69-100   \\
$\Delta(1700)3/2^-$  & 1.704 $\pm$ 0.008 &  0.295 $\pm$ 0.035  &   79-100   \\
$\Delta(1905)5/2^+$  & 1.883 $\pm$ 0.019 &  0.327 $\pm$ 0.069  &   70-100   \\
$\Delta(1950)7/2^+$  & 1.943 $\pm$ 0.018  & 0.230 $\pm$ 0.088  &   37-77   \\ \hline
\end{tabular}
\caption{Resonance masses, total decay widths, and branching fractions to the $\pi\pi N$ final states determined
from the $\pi^+\pi^-p$ photoproduction data for the excited nucleon states listed in Table~\ref{nstpar}.}
\label{hadrpar} 
\end{center}
\end{table*}

\section{Summary} 
\label{summ}

The first results on nine 1-fold differential and fully integrated $\pi^+\pi^-p$ photoproduction cross sections off the
proton in the range of $W$ from 1.6~GeV to 2.0~GeV have become available from measurements with the CLAS detector at 
Jefferson Lab. These data amount to a factor of $\sim$50 increase in the number of events from this reaction compared to
previous measurements. Analysis of these cross sections of much improved accuracy has allowed us, by using the updated
JM17 meson-baryon reaction model, to establish all essential contributing mechanisms to the process from their
manifestations in the  observables and to extract the resonant contributions to the experimental data. The good description
of the experimental data achieved in the entire $W$ range provides confidence in the procedure we have used to
determine the resonant contributions to the differential cross sections from the data fit.

Using a unitarized Breit-Wigner ansatz~\cite{Mo12,Ait72}, which allowed us to account for the restrictions imposed by a general 
unitarity condition on the resonant amplitudes, the resonance photocouplings were determined from the resonance contributions.  
For the first time, the nucleon resonance photocouplings for the states in the mass range from 1.6~GeV to 2.0~GeV were 
determined from the analysis of the data on $\pi^+\pi^-p$ photoproduction. The $\Delta(1620)1/2^-$, $\Delta(1700)3/2^-$, 
$N(1720)3/2^+$, and $\Delta(1905)5/2^+$ resonance photocouplings were extracted from the $\pi^+\pi^-p$ photoproduction 
channel with much improved  accuracy compared to previous $\pi N$ analyses, because of the preferential decays of
these resonances to the $\pi \pi N$ final states with branching fractions above 70\%. The results on $\pi\pi N$
photoproduction from this work and multichannel analyses \cite{So17,BnGa12,BnGa14} are now the major source of
information on the photocouplings of these states. The results on the $N^*$ photocouplings from $\pi^+\pi^-p$ photoproduction
show good consistency with the ranges for the photocouplings from the PDG listings~\cite{Rpp18}, which is an important result
considering the much larger cross sections of this channel in comparison with the $\pi^0\pi^0p$ channel, which were analyzed
so far within the $W$ range of our measurements~\cite{Tho05}. 
Implementation of our data into
the coupled channel analyses will help to check further the extraction of the resonance photocouplings within the JM17 model.
The results presented in this paper pave the way for the future combined analysis of the $\pi^+\pi^-p$ photo- and
electroproduction data from CLAS, which has already revealed substantial evidence for the new $N'(1720)3/2^+$ baryon
state~\cite{Mo16b}.

\section{Acknowledgments}

We would like to acknowledge the outstanding efforts of the staff of the Accelerator 
and the Physics Divisions at Jefferson Lab 
that made this experiment possible. 
This work was supported in part by
the Chilean Comisi\'on Nacional de Investigaci\'on Cient\'ifica y 
Tecnol\'ogica (CONICYT),
the Italian Istituto Nazionale di Fisica Nucleare,
the French Centre National de la Recherche Scientifique,
the French Commissariat \`{a} l'Energie Atomique,
the U.S. Department of Energy,
the National Science Foundation,
the Scottish Universities Physics Alliance (SUPA),
Skobeltsyn Institute of Nuclear Physics,
the Physics Department at Moscow State University, 
Ohio University,
the University of South Carolina,
the United Kingdom's Science and Technology Facilities Council (STFC),
and the National Research Foundation of Korea.
Authored by Jefferson Science Associates, LLC under U.S. DOE Contract No. DE-AC05-06OR23177. The U.S.
Government retains a non-exclusive, paid-up, irrevocable, world-wide license to publish or reproduce this
manuscript for U.S. Government purposes. This material is based upon work supported by the U.S. Department
of Energy, Office of Science, Office of Nuclear Physics under contract DE-AC05-06OR23177.




\begin{thebibliography}{99}

\bibitem{Az13} I.G. Aznauryan {\it et al.}, Int. J. Mod. Phys. {\bf E22}, 1330015 (2013).

\bibitem{Du12} J.J. Dudek and R.G. Edwards, Phys. Rev. D {\bf 85}, 054016 (2012).

\bibitem{Ro11} H.L.L. Roberts {\it et al.}, Few Body Syst. {\bf 51}, 1 (2011).

\bibitem{Cr13} V. Crede and W. Roberts, Rep. Prog. Phys. {\bf 76}, 076301 (2013).

\bibitem{Kl17} E. Klempt {\it et al.}, EPJ Web Conf. {\bf 134}, 02002 (2017). 

\bibitem{Be17} R. Beck {\it et al.}, EPJ Web Conf. {\bf 134}, 02001 (2017).

\bibitem{So17} V. Sokhoyan {\it et al.}, Eur. Phys. J. A{\bf 51}, 95 (2015).



\bibitem{Leps} H. Kori, JPS Conf. Proc. {\bf 10}, 01008 (2016).

\bibitem{Gu14} E. Gutz {\it et al.} {\it (CBELSA/TAPS Collaboration)}, Eur. Phys. J. A{\bf 50}, 74 (2014).

\bibitem{Ass03} Y. Assafiri {\it et al.}, Phys. Rev. Lett. {\bf 90}, 222001 (2003).

\bibitem{Kash12} V.L. Kashevarov {\it et al.}, Phys. Rev. C {\bf 85}, 064610 (2012).

\bibitem{Ann15} J.R.M. Annand {\it et al.} {\it (A2 Collaboration)}, Phys. Rev. C {\bf 91}, 055208 (2015).

\bibitem{Bu16} V.D. Burkert, Few Body Syst. {\bf 57}, 873 (2016).

\bibitem{BnGa12} A.V. Anisovich {\it et al.}, Eur. Phys. J. A{\bf 48}, 15 (2012).

\bibitem{BnGa14} A.V. Anisovich {\it et al.}, Eur. Phys. J. A{\bf 50}, 129 (2014).

\bibitem{An16} A.V. Anisovich {\it et al.}, Eur. Phys. J. {\bf A52}, 284 (2016). 

\bibitem{Rpp18} M. Tanabashi {\it et al.} (Particle Data Group), Phys. ReV. D {\bf 98}, 03001 (2018).

\bibitem{Bu17} A.V. Anisovich {\it et al.}, Eur. Phys. J. A{\bf 53}, 242 (2017).

\bibitem{Brad06} R.K. Bradford {\it et al.}  Phys. Rev. C {\bf 73}, 035202 (2006).

\bibitem{Brad07} R.K. Bradford {\it et al.} {\it (CLAS Collaboration)}, Phys. Rev. C {\bf 75}, 035205 (2007).

\bibitem{McC09} M.E. McCracken {\it et al.} {\it (CLAS Collaboration)}, Phys. Rev. C {\bf 81}, 025201 (2010).

\bibitem{Dey10} B. Dey {\it et al.} {\it (CLAS Collaboration)}, Phys. Rev. C {\bf 82}, 025202 (2010).

\bibitem{Capst} S. Capstick and W. Roberts, Prog. Part. Nucl. Phys. {\bf 45}, S241 (2000).

\bibitem{Mo16b} V.I. Mokeev {\it et al.}, EPJ Web Conf. {\bf 113}, 01013 (2016).

\bibitem{Ri03} M. Ripani {\it et al.} {\it (CLAS Collaboration)}, Phys. Rev. Lett. {\bf 91}, 022002 (2003).

\bibitem{Is16} E.L. Isupov {\it et al.} {\it (CLAS Collaboration)}, Phys. Rev. C {\bf 96}, 025209 (2017).

\bibitem{Tho05} U. Thoma, Int. J. Mod. Phys. A {\bf 20}, 280 (2005).

\bibitem{Mo12} V.I. Mokeev {\it et al.} {\it (CLAS Collaboration)}, Phys. Rev. C {\bf 86}, 035203 (2012).

\bibitem{Mo16a} V.I. Mokeev {\it et al.}, Phys. Rev. C {\bf 93}, 054016 (2016).

\bibitem{Mo09} V.I. Mokeev {\it et al.}, Phys. Rev. C {\bf 80}, 045212 (2009).

\bibitem{Mo16} V.I. Mokeev, Few Body Syst. {\bf 57}, 909 (2016). 

\bibitem{Me03} B.A. Mecking {\it et al.}, Nucl. Instr. and Meth. {\bf A503}, 513 (2003).

\bibitem{tagger} D. I. Sober {\it et al.}, Nucl. Instrum. Meth. A {\bf 440}, 263 (2000).

\bibitem{bradnote} R. Bradford and R.A. Schumacher. CLAS-Note 2002-003,\\
\url{https://www.jlab.org/Hall-B/notes/clas_notes02/02-003.pdf}.

\bibitem{willthesis} M. Williams, Carnegie Mellon University Ph.D. Thesis, (2007).

\bibitem{dcs} M.D. Mestayer {\it et al.}, Nucl. Instr. and Meth. {\bf A449}, 81 (2000).

\bibitem{tof} E. Smith {\it et al.}, Nucl. Instr. and Meth. {\bf A432}, 265 (1999).

\bibitem{stt} Y. Sharabian {\it et al.}, Nucl. Instr. and Meth. {\bf A559}, 246 (2006).

\bibitem{eloss} E. Pasyuk, CLAS-Note 2007-016,\\
\url{https://misportal.jlab.org/ul/Physics/Hall-B/clas/viewFile.cfm/2007-016.pdf?documentId=423}.

\bibitem{St07} S. Stepanyan {\it et al.}, Nucl. Instrum. Meth. A {\bf 572}, 654 (2000).

\bibitem{kinfit} M. Williams and C.A. Meyer, CLAS-Note 2003-017,\\
\url{https://www.jlab.org/Hall-B/notes/clas_notes03/03-017.pdf}.

\bibitem{saphir} C. Wu {\it et al.}, Eur. Phys. J. A {\bf 23}, 317 (2005). 

\bibitem{Fe09} G.V. Fedotov {\it et al.} {\it (CLAS Collaboration)}, Phys. Rev. C {\bf 79}, 015204 (2009).

\bibitem{gsim} M. Holtrop, \url{http://nuclear.unh.edu/~maurik/gsim_info.shtml}.

\bibitem{Ri00} M. Ripani {\it et al.}, Nucl. Phys. A {\bf 672}, 220 (2000). 

\bibitem{Mo05} I.G. Aznauryan {\it et al.}, Phys. Rev. C {\bf 72}, 045201 (2005). 

\bibitem{flux} J.~Ball and E.~Pasyuk, CLAS-Note 2005-002,\\
\url{https://misportal.jlab.org/ul/Physics/Hall-B/clas/viewFile.cfm/2005-002.pdf?documentId=24}.

\bibitem{abbhhm} ABBHHM Collaboration, Phys. Rev. {\bf 188}, 2060 (1969). 

\bibitem{clasdb} CLAS Physics Database \url{https://clasweb.jlab.org/physicsdb/}.

\bibitem{Ait72} I.J.R. Aitchison, Nucl. Phys. A {\bf 189}, 417 (1972).

\bibitem{Man92} D.M. Manley and E.M. Salesky, Phys. Rev. D {\bf 45}, 4002 (1992). 

\bibitem{Dug09} M. Dugger {\it et al.}  {\it (CLAS Collaboration)}, Phys. Rev. C {\bf 79}, 065206 (2009).

\bibitem{Mo14} V.I. Mokeev and I.G. Aznauryan, Int. J. Mod. Phys. Conf. Ser. {\bf 26}, 146080 (2014).





\end{thebibliography}
\end{document}